\documentclass[12pt,a4paper]{article}
\usepackage{epsfig}

\usepackage{graphics}

\usepackage{graphicx}

\usepackage{dcolumn}

\usepackage{bm}
\newcommand{\sba}{\begin{subeqnarray}}
\newcommand{\sea}{\end{subeqnarray}}

\begin{document}
\title{Diagrammatic theory for twofold degenerate Anderson impurity model}
\author{V.\ A.\ Moskalenko$^{1,2}$, L.\ A.\ Dohotaru$^{3}$, D.\ F.\ Digor$^{1}$ and I.\ D.\ Cebotari$^{1}$}
\maketitle {$^{1}$Institute of Applied Physics, Moldova Academy of
Sciences, Chisinau 2028, Moldova,\\$^{2}$BLTP, Joint Institute for
Nuclear Research, 141980 Dubna, Russia,\\ $^{3}$ Technical
University, Chisinau 2004, Moldova}
\begin{abstract}
The twofold degenerate Anderson impurity model [1-4] is investigated
and the strong electronic correlations of d-electrons of impurity
ion are taken into account by elaborating suitable diagram
technique.

We discuss the properties of the Slater-Kanamori model [2-4] of
d-impurity electrons. After finding the eigenfunctions and
eigenvalues of all 16 local states, we determine the local
one-particle propagator. Then we construct the perturbation theory
around the atomic limit of the impurity ion and obtain the Dyson
type equation for the renormalized  one-particle propagator.
Diagrammatic theory has been developed and correlation function
determined. Special diagrammatic approximation was discussed and
summation of diagram has been considered.
\end{abstract}

PACS numbers: {71.27.+a, 71.10.Fd}

\section{Introduction}

The theory of strongly correlated electron systems plays a central
role in contemporary condensed matter physics. The essence of the
problem is the competition between the localization tendency
originated by the Coulomb repulsion of d electrons and itinerancy
tendency arising as a result of hybridization of electron orbitals.

The orbital degeneracy can be completely eliminated in solid
substances but in many of them, for example, new superconductors
based on $Fe$ and $AnC_{60}$ materials orbital degeneracy is not
completely eliminated and orbital effects are important. For
instance, orbital degeneracy plays essential role in the Mott
metal-insulator transition. Here the effects of Hund's rule coupling
in our orbitally degenerated model are studied with diagrammatic
approach.

We study the influence of the intra-atomic Coulomb interactions of
the two electrons with opposite spins situated on the same or
different orbitals and intra-atomic exchange is analyzed.

Our investigation is based on the diagram theory elaborated for
strongly correlated electron systems as in non-degenerated
[5-9,11-14]and as in twofold degenerated ones [10].

The paper has the following structure. In Sec. 1 we describe the
twofold degenerate Anderson impurity model. The local properties of
our model are considered in Sec. 2. The perturbation theory around
the atomic limit of impurity ion is formulated in Sec. 3. In this
section we discuss the process of delocalization and renormalization
of the dynamical quantities. In Sec. 4 the simplest irreducible
Green's function is calculated. Sec. 5 is devoted to discussion of
the Mott-Hubbard phase transition. Sec. 6 is devoted to the
conclusions.

The Anderson Impurity model with twofold orbital degeneracy has the
Hamiltonian composed one part of conduction electrons - one part of
interacting localized and strongly correlated electrons and of
hybridization term between these two parts [1-4]:

\begin{eqnarray}
H &=& H^{0} + H_{int}, \label{1} \\
H^{0}&=& H_{c}^{0}+H_{d}^{L}, \label{2} \\
H_{c}^{0}&=&\sum\limits_{\vec{k}l\sigma}\epsilon_{l}(\vec{k})
C_{\vec{k}l\sigma}^{+}C_{\vec{k}l\sigma},
\label{3} \\
H_{d}^{L}&=&\sum\limits_{l,\sigma}\epsilon^{d}d_{l\sigma}^{+}d_{l\sigma}
+U\sum\limits_{l}n_{l\uparrow}n_{l\downarrow}
+U^{\prime}n_{1}n_{2}+I_{H}\sum\limits_{\sigma\sigma^{\prime}}
d_{1\sigma}^{+}d_{2\sigma}^{+}d_{1\sigma^{\prime}}d_{2\sigma^{\prime}}\\
\nonumber
&+&I_{H}^{\prime}(d_{1\uparrow}^{+}d_{1\downarrow}^{+}d_{2\downarrow}d_{2\uparrow}+H.C.),\\
H_{int}&=&\frac{1}{\sqrt{N}}\sum\limits_{\vec{k}l\sigma}(V_{\vec{k}l}d_{l\sigma}^{+}
C_{\vec{k}l\sigma}+V_{\vec{k}l}^{*}C_{\vec{k}l\sigma}^{+}d_{l\sigma}),\label{1-5}
\end{eqnarray}%
where the local Hamiltonian $H_{d}^{L}$ is standard Slater-Kanamori
[2-4] form, $C_{\vec{k}l\sigma}$ is conduction electron annihilation
operator with momentum $\vec{k}$, orbital number $l=1,2$ and spin
$\sigma=\pm 1(\uparrow ,\downarrow)$, $d_{l\sigma}$ operator for
localized $d$ electron. Conduction electron of $l-th$ orbital state
hybridizes only with the local electron of the same orbital state.
$n_{l\sigma}=d_{l\sigma}^{+}d_{l\sigma}$,
$n_{l}=\sum\limits_{\sigma}n_{l\sigma}$, $V_{\vec{k}l}$ is matrix
element of hybridization. $U$ is Coulomb repulsion between the
d-electrons in the same orbital state and $U^{\prime}$ - between
electrons in different orbital states. $I_{H}$ is coefficient of the
Hund's rule coupling and pair hopping terms, $\epsilon_{l}(\vec{k})$
is the band dispersion and $\epsilon^{d}$ - is impurity ion energy
evaluated from the chemical potential $\mu$. $N$ is a number of
lattice sites.

In the following we assume that the symmetry of the system is such
that exist the relation:

\begin{eqnarray}
U^{\prime}=U-2I_{H},    I_{H}^{\prime}=I_{H}.\label{6}
\end{eqnarray}%
%
The Coulomb interactions are far too large to be treated as
perturbation and they must be included in $H^{0}$ - zero order
Hamiltonian. The hybridization term (5) is considered as the
perturbation of the system. In the following the main ideas of the
perturbation theory elaborated for non-degenerate strongly
correlated systems are extended for degenerated systems. Such
generalization has been discussed, for example, in the case of
twofold degenerate Hubbard model [10]. As is known the new elements
of this perturbation theory of strongly correlated systems are the
irreducible correlation functions which contain all charge, spin and
pairing quantum fluctuations.

\section{Local properties}

%
In the main approximation of the Anderson model one has free
conduction and strongly interacting localized electrons described by
the Hamiltonian $H_{0}$. The localized part of the Hamiltonian,
$H_{0}^{L}$, can be diagonalized by using Hubbard transfer operators
$\chi^{mn}=|m\rangle\langle n|$ where $|m\rangle$ is eigenvector of
operator $H_{d}^{L}$ [5].

Because orbital quantum number takes two values $l=1,2$ the total
number of local quantum states is equal to 16.

There are the following eigenvectors of operator $H_{d}^{L}$. The
first quantum state $|1\rangle$ is the vacuum state $|0\rangle$ with
energy $E_{1}=0$. There are 4 one particle states with spin
$S=\frac{1}{2}$ and $S_{z}=\pm\frac{1}{2}$:\\
$|2\rangle=d_{1\uparrow}^{+}|0\rangle$,
$|3\rangle=d_{2\uparrow}^{+}|0\rangle$,
$|4\rangle=d_{1\downarrow}^{+}|0\rangle$ and
$|5\rangle=d_{2\downarrow}^{+}|0\rangle$. The energies of all these
states are $E_{2}=E_{3}=E_{4}=E_{5}=\epsilon_{d}$.

Then there are six states with two particles. Three of them are
singlet states with spin $S=0$ and others 3 triplet states with
$S=1$ and $S_{z}=-1,0,1$,

$|6\rangle =
\frac{1}{\sqrt{2}}(d_{1\uparrow}^{+}d_{1\downarrow}^{+}-d_{2\uparrow}^{+}d_{2\downarrow}^{+})|0\rangle$,
$|7\rangle =
\frac{1}{\sqrt{2}}(d_{1\uparrow}^{+}d_{1\downarrow}^{+}+d_{2\uparrow}^{+}d_{2\downarrow}^{+})|0\rangle$,\\
$|8\rangle =
\frac{1}{\sqrt{2}}(d_{1\uparrow}^{+}d_{2\downarrow}^{+}-d_{1\downarrow}^{+}d_{2\uparrow}^{+})|0\rangle$,
$|9\rangle =d_{1\uparrow}^{+}d_{2\uparrow}^{+}|0\rangle$,\\
$|10\rangle =
\frac{1}{\sqrt{2}}(d_{1\uparrow}^{+}d_{2\downarrow}^{+}+d_{1\downarrow}^{+}d_{2\uparrow}^{+})|0\rangle$,
$|11\rangle = d_{1\downarrow}^{+}d_{2\downarrow}^{+}|0\rangle$.\\
The eigenvalues of these quantum states are

$E_{6}=2\epsilon_{d}+U-I_{H}^{\prime}$, $E_{7}=2\epsilon_{d}+U+I_{H}^{\prime}$, $E_{8}=2\epsilon_{d}+U^{\prime}+I_{H}$,\\
$E_{9}=E_{10}=E_{11}=2\epsilon_{d}+U^{\prime}-I_{H}$.

Then there are four states composed from three particles

$|12\rangle=d_{1\uparrow}^{+}d_{1\downarrow}^{+}d_{2\uparrow}^{+}|0\rangle$,
$|13\rangle=d_{2\uparrow}^{+}d_{2\downarrow}^{+}d_{1\uparrow}^{+}|0\rangle$,\\
$|14\rangle=d_{1\uparrow}^{+}d_{1\downarrow}^{+}d_{2\downarrow}^{+}|0\rangle$,
$|15\rangle=d_{2\uparrow}^{+}d_{2\downarrow}^{+}d_{1\downarrow}^{+}|0\rangle$,\\
with energy value
$E_{12}=E_{13}=E_{14}=E_{15}=3\epsilon_{d}+U+2U^{\prime}-I_{H}$.

The last local state is singlet

$|16\rangle=d_{1\uparrow}^{+}d_{1\downarrow}^{+}d_{2\uparrow}^{+}d_{2\downarrow}^{+}|0\rangle$
with energy value $E_{16}=4\epsilon_{d}+2U+4U^{\prime}-2I_{H}$.

When equalities $(6)$ take place we obtain more simple forms:

$E_{6}=E_{8}=2\epsilon_{d}+U-I_{H}$, $E_{7}=2\epsilon_{d}+U+I_{H}$,
$E_{9}=2\epsilon_{d}+U-3I_{H}$, $E_{12}=3\epsilon_{d}+3U-5I_{H}$,
 $E_{16}=4\epsilon_{d}+6U-10I_{H}$.

The triplet states $|9\rangle$, $|10\rangle$ and $|11\rangle$ are
the lowest by energy.

Quantum states enumerated above permit us to organize Hubbard
transfer operators $\chi^{mn}$  and establish the relation with
fermion impurity operators [10]:
\begin{eqnarray}
d_{l\sigma}^{+}=\chi^{2+l-\sigma,1}+\frac{\sigma}
{\sqrt{2}}[(-1)^{l+1}\chi^{6,2+l+\sigma}+\chi^{7,2+l+\sigma}]+
\frac{1}{\sqrt{2}}[\sigma \chi^{8,5-l+\sigma}
+(-1)^{l+1}\chi^{10,5-l+\sigma} ] \\ \nonumber
+\frac{1}{\sqrt{2}}[-\chi^{12+l-\sigma,8}
+\sigma(-1)^{l+1}\chi^{12+l-\sigma,10} ]
+\frac{1}{\sqrt{2}}[-(1)^{l}\chi^{15-l-\sigma,6}
+\chi^{15-l-\sigma,7} ]+
\\ \nonumber(-1)^{l+1}X^{10-\sigma,5-l-\sigma}+
(-1)^{l+1}\sigma\chi^{12+l+\sigma,10+\sigma}+ \sigma \chi^{16,
15-l+\sigma}. \label{7}
\end{eqnarray}

Equation $(7)$ allows to calculate all the local dynamical
quantities. For example quantum electron number has the form:
\begin{eqnarray}
n_{l\sigma}= \chi^{2+l-\sigma,2+l-\sigma}+ \frac{1}{2}[
\chi^{6,6}+(-1)^{l+1}\chi^{6,7} + (-1)^{l+1}\chi^{7,6}+X^{7,7}] \\
\nonumber +\frac{1}{2}[ \chi^{8,8}+\sigma(-1)^{l+1}\chi^{8,10}+
\sigma(-1)^{l+1}\chi^{10,8}+\chi^{10,10} ]  \\ \nonumber
\chi^{10-\sigma, 10-\sigma}+\chi^{12+l-\sigma, 12+l-\sigma}+
\chi^{12+l+\sigma, 12+l+\sigma} \\ \nonumber +\chi^{15-l-\sigma,
15-l-\sigma}+\chi^{16,16}, \label{8}
\end{eqnarray}%
and
\begin{equation}
n_{l\uparrow}-n_{l\downarrow}= \chi^{1+l, 1+l}- \chi^{3+l, 3+l}+
(-1)^{l+1}[\chi^{8,10}+\chi^{10,8}] + \chi^{9,9}-
\chi^{11,11}+\chi^{14-l, 14-l}-\chi^{16-l, 16-l}. \label{9}
\end{equation}

For $\tau$ dependent quantity $A(\tau)=e^{\tau H_{0}}Ae^{-\tau
H_{0}}$ we have the equation:
\begin{eqnarray}
n_{l\uparrow}(\tau) -n_{l\downarrow}(\tau)= \chi^{1+l, 1+l}-
\chi^{3+l, 3+l}+ (-1)^{l+1} [\chi^{8, 10}e^{\tau (E_{8}-E_{10})} \\
\nonumber +\chi^{10,8}e^{\tau(E_{10}-E{8})}]
+\chi^{9,9}-\chi^{11,11}+ \chi^{14-l,14-l}- \chi^{16-l,
16-l}.\label{10}
\end{eqnarray}%
%
%
The correlation between quantities with different orbital numbers is
determined by the equation $(l=1,2)$:
\begin{eqnarray}
(n_{l\uparrow}(\tau)-n_{l\downarrow}(\tau))
(n_{l^{\prime}\uparrow}(0)-n_{l^{\prime}\downarrow}(0))=
\delta_{ll^{\prime}} [\chi^{1+l, 1+l}+\chi^{3+l, 3+l} +\chi^{14-l,
14-l}+ \\ \nonumber \chi^{16-l, 16-l} ] + (-1)^{l+l^{\prime}}[
\chi^{8,8}e^{\tau(E_{8}-E_{10})} +\chi^{10,
10}e^{\tau(E_{10}-E_{8})}]+ \chi^{9, 9}+\chi^{11, 11},   \label{11}
\end{eqnarray}
\begin{eqnarray}
\sum\limits_{ll^{\prime}}(n_{l\uparrow}(\tau)-n_{l\downarrow}(\tau))
(n_{l^{\prime}\uparrow}(0)-n_{l^{\prime}\downarrow}(0))=
\frac{4}{Z_{0}}(e^{-\beta E_{2}}+e^{-\beta E_{12}}+2e^{-\beta
E_{9}}). \label{12}
\end{eqnarray}
In special case $l=1, l^{\prime}=2$ we have:
\begin{eqnarray}
(n_{1\uparrow}(\tau)-n_{1\downarrow}(\tau))
(n_{2\uparrow}(0)-n_{2\downarrow}(0))= -[ \chi^{8,8}e^{ \tau(
E_{8}-E_{10}) } + \chi^{10,10}e^{\tau(E_{10}-E_{8})} ] +
\chi^{9,9}+\chi^{11,11},  \label{13}
\end{eqnarray}
which is the $d$ electron susceptibility [2-4].

We now define the Matsubara one-particle Green's function of
localized d-electrons:

\begin{eqnarray}
g^{0}(l\sigma\tau,l^{\prime}\sigma^{\prime}\tau^{\prime})=
g^{0}_{l\sigma,l^{\prime}\sigma^{\prime}}(\tau-\tau^{\prime})=
-\langle T
d_{l\sigma}(\tau)\bar{d}_{l^{\prime}\sigma^{\prime}}(\tau^{\prime})
\rangle_{0}, \label{14}
\end{eqnarray}%
where $d_{l\sigma}(\tau)= e^{\tau H_{0}}d_{l\sigma}e^{-\tau H_{0}}$,
 $\bar{d_{l\sigma}}(\tau)=e^{\tau H_{0}}d_{l\sigma}^{+}e^{-\tau
 H_{0}}$.

The Fourier components of this Green's function are:

\begin{eqnarray}
g^{(0)}(\tau)= \frac{1}{\beta}
\sum\limits_{\omega_{n}}e^{-i\omega_{n}\tau} g^{(0)}(i\omega_{n}).
\label{15}
\end{eqnarray}%

Using $(8)$ and the properties of Hubbard operators we obtain the
equation for local function:
\begin{eqnarray}
g_{l\sigma l^{\prime}\sigma^{\prime}}^{(0)}(i\omega_{n}) =
\frac{\delta_{ll^{\prime}}\delta_{\sigma\sigma^{\prime}}}{Z_{0}} \{
\frac{e^{-\beta E_{1}}+e^{-\beta E_{2}}} {i\omega_{n}+E_{1}-E_{2}} +
\frac{e^{-\beta E_{2}}+e^{-\beta E_{6}}} {i\omega_{n}+E_{2}-E_{6}} +
\\ \nonumber \frac{1}{2} \frac{e^{-\beta E_{2}}+e^{-\beta E_{7}}}
{i\omega_{n}+E_{2}-E_{7}} + \frac{3}{2} \frac{e^{-\beta
E_{2}}+e^{-\beta E_{9}}} {i\omega_{n}+E_{2}-E_{9}} + \frac{e^{-\beta
E_{6}}+e^{-\beta E_{12}}} {i\omega_{n}+E_{6}-E_{12}} + \\ \nonumber
\frac{1}{2} \frac{e^{-\beta E_{7}}+e^{-\beta E_{12}}}
{i\omega_{n}+E_{7}-E_{12}} + \frac{3}{2} \frac{e^{-\beta
E_{9}}+e^{-\beta E_{12}}} {i\omega_{n}+E_{9}-E_{12}} +
\frac{e^{-\beta E_{12}}+e^{-\beta E_{16}}}
{i\omega_{n}+E_{12}-E_{16}} \label{16} \},
\end{eqnarray}
where $Z_{0}$ is partition function in atomic limit
\begin{equation}
Z_{0}=e^{-\beta E_{1}}+4 e^{-\beta E_{2}}+2 e^{-\beta E_{6}}+
e^{-\beta E_{7}}+3 e^{-\beta E_{9}}+4 e^{-\beta E_{12}}+e^{-\beta
E_{16}}.\label{17}
\end{equation}

The spectral function of impurity d-electron in local approximation
is equal to
\begin{equation}
A^{(0)}(E) = - 2Img^{0}(E+i\delta), \label{18}
\end{equation}
where $g^{0}(E+i\delta )$ with $\delta=+0$ is analytical
continuation of the Matsubara to retarded Green's function.

Using $(14)$ we obtain
\begin{eqnarray}
A^{(0)}(E)=\frac{2\pi}{Z_{0}} \{ (e^{-\beta E_{1}}+ e^{-\beta
E_{2}})\delta (E+ E_{1}-E_{2})+ (e^{-\beta E_{2}}+ e^{-\beta
E_{6}})\delta (E+ E_{2}-E_{6})+\nonumber\\ \frac{1}{2} (e^{-\beta
E_{2}}+ e^{-\beta E_{7}})\delta (E+ E_{2}-E_{7})+ \frac{3}{2}
(e^{-\beta E_{2}}+ e^{-\beta E_{9}})\delta (E+ E_{2}-E_{9})+\\
(e^{-\beta E_{6}}+ e^{-\beta E_{12}})\delta (E+ E_{6}-E_{12})+
\frac{1}{2} (e^{-\beta E_{7}}+ e^{-\beta E_{12}})\delta
(E+E_{7}-E_{12})+\nonumber \\ \frac{3}{2} (e^{-\beta E_{9}}+
e^{-\beta E_{12}})\delta (E+ E_{9}-E_{12})+(e^{-\beta E_{12}}+
e^{-\beta E_{16}})\delta (E+ E_{12}-E_{16}),\nonumber \label{19}
\end{eqnarray}
with property
\begin{equation}
\int_\infty^\infty A^{(0)}(E) dE= 2 \pi. \label{20}
\end{equation}

\section{Delocalization processes}

We use the perturbation theory elaborated previously for strongly
correlated electron systems both of non degenerate [5-9,11-14] and
of degenerate forms [10]. We study the process of renormalization of
Green's function resulting from intra- and inter-orbital flips of
tunneling electrons.

The full Matsubara Green's function in the interaction
representation for conduction and impurity electrons are:

\begin{eqnarray}
G(\vec{k}l\sigma\tau|\vec{k^{\prime}}l^{\prime}\sigma^{\prime}\tau^{\prime})
= -\langle T
C_{\vec{k}l\sigma}(\tau)\bar{C}_{\vec{k}^{\prime}l^{\prime}\sigma^{\prime}}(\tau^{\prime})
U(\beta) \rangle_{0}^{c}, \\ \nonumber
g(l\sigma\tau|l^{\prime}\sigma^{\prime}\tau^{\prime}) = -\langle T
d_{l\sigma}(\tau)\bar{d}_{l^{\prime}\sigma^{\prime}}(\tau^{\prime})
U(\beta) \rangle_{0}^{c}. \label{21}
\end{eqnarray}

The anomalous functions are defined as

\begin{eqnarray}
F(\vec{k}l\sigma\tau|\vec{k}^{\prime}l^{\prime}\sigma^{\prime}\tau^{\prime})
&=& -\langle T
C_{\vec{k}l\sigma}(\tau)C_{\vec{k}^{\prime}l^{\prime}\sigma^{\prime}}(\tau^{\prime})
U(\beta) \rangle_{0}^{c}, \\ \nonumber
\bar{F}(\vec{k}l\sigma\tau|\vec{k}^{\prime}l^{\prime}\sigma^{\prime}\tau^{\prime})
&=& -\langle T
\bar{C}_{\vec{k}l\sigma}(\tau)\bar{C}_{\vec{k}^{\prime}l^{\prime}\sigma^{\prime}}(\tau^{\prime})
U(\beta) \rangle_{0}^{c}, \\ \nonumber
f(l\sigma\tau|l^{\prime}\sigma^{\prime}\tau^{\prime}) &=& -\langle T
d_{l\sigma}(\tau)d_{l^{\prime}\sigma^{\prime}}(\tau^{\prime})
U(\beta) \rangle_{0}^{c}, \\ \nonumber
\bar{f}(l\sigma\tau|l^{\prime}\sigma^{\prime}\tau^{\prime})&=&
-\langle
T\bar{d}_{l\sigma}(\tau)\bar{d}_{l^{\prime}\sigma^{\prime}}(\tau^{\prime})
U(\beta) \rangle_{0}^{c}. \label{22}
\end{eqnarray}

Here $\tau$ and $\tau^{\prime}$ stand for imaginary time with $0 \le
\tau \le \beta$, $\beta$ is inverse temperature, $T$ is
chronological ordering operator.

The evolution operator is
\begin{equation}
U(\beta) = T \exp{(-\int_0^{\beta}  H_{int}(\tau)  d
\tau)}.\label{23}
\end{equation}

The statistical averaging is carried out in $(21)$ and $(22)$ with
respect to the zero-order density matrix of the conduction and
impurity electrons. Index $c$ means connected diagrams.

In the zero order approximation we have
\begin{eqnarray}
H_{0}^{L}&=& \sum\limits_{n=1}^{16} E_{n}\chi^{nn}, \,\ \,\
 \sum\limits_{n=1}^{16}\chi^{nn}=1, \\ \nonumber
G_{l\sigma
l^{\prime}\sigma^{\prime}}^{(0)}(\vec{k}\vec{k^{\prime}}|\tau-\tau^{\prime})&=&
\delta_{\vec{k}\vec{k}^{\prime}}\delta_{ll^{\prime}}\delta_{\sigma\sigma^{\prime}}
G_{l\sigma}^{(0)}(\vec{k}|\tau-\tau^{\prime}), \\ \nonumber
G_{l\sigma}^{(0)}(\vec{k}|\mathrm{i}\omega_{n})&=&
\frac{1}{\mathrm{i}\omega_{n}-\epsilon(\vec{k})},
\omega_{n}=\frac{(2n+1)\pi}{\beta}.\label{24}
\end{eqnarray}
and $g^{(0)}(\mathrm{i}\omega_{n})$ is determined by the equation
$(16)$.

Hybridization between the conduction and $d$ impurity electrons
results in renormalization of their propagators. Because the number
of conduction electrons $N$ is much larger than the single impurity
state, the effect of the latter on the conduction band scales as
$\frac{1}{N}.$

The renormalized conduction electron propagator is
\begin{eqnarray}
G_{l\sigma
l^{\prime}\sigma^{\prime}}(\vec{k}\vec{k^{\prime}}|\mathrm{i}\omega_{n})=
\delta_{\vec{k}\vec{k}^{\prime}}\delta_{ll^{\prime}}\delta_{\sigma\sigma^{\prime}}
G_{l\sigma}^{(0)}(\vec{k}|\mathrm{i}\omega_{n})+ \\ \nonumber
\frac{V_{\vec{k}l}^*V_{\vec{k}^{\prime}l^{\prime}}}{N}
G_{l\sigma}^{(0)}(\vec{k}|\mathrm{i}\omega_{n})
g_{l\sigma,l^{\prime}\sigma^{\prime}}(\mathrm{i}\omega_{n})
G_{l^{\prime}\sigma^{\prime}}^{(0)}(\vec{k}^{\prime}|\mathrm{i}\omega_{n}),
\label{25}
\end{eqnarray}
where $g_{l\sigma l^{\prime}\sigma^{\prime}}(\mathrm{i}\omega_{n})$
is the full impurity electron propagator.

A similar equation holds for the anomalous function of conduction
electrons in superconducting state:
\begin{eqnarray}
F_{l\sigma
l^{\prime}\sigma^{\prime}}(\vec{k},-\vec{k^{\prime}}|\mathrm{i}\omega_{n})=
\frac{V_{\vec{k}l}^*V_{\vec{k}^{\prime}l^{\prime}}}{N}
G_{l\sigma}^{(0)}(\vec{k}|\mathrm{i}\omega_{n}) f_{l\sigma
l^{\prime}\sigma^{\prime}}(\mathrm{i}\omega_{n})
G_{l^{\prime}\sigma^{\prime}}^{(0)}(-\vec{k}^{\prime}|-\mathrm{i}\omega_{n}).
\nonumber\label{0}
\end{eqnarray}
The equations for the full functions $g$ and $f$ of impurity
electrons have the diagrammatical form shown in Fig.1.
%
\begin{figure*}[h]
%
\centering
\includegraphics[width=0.85\textwidth,clip]{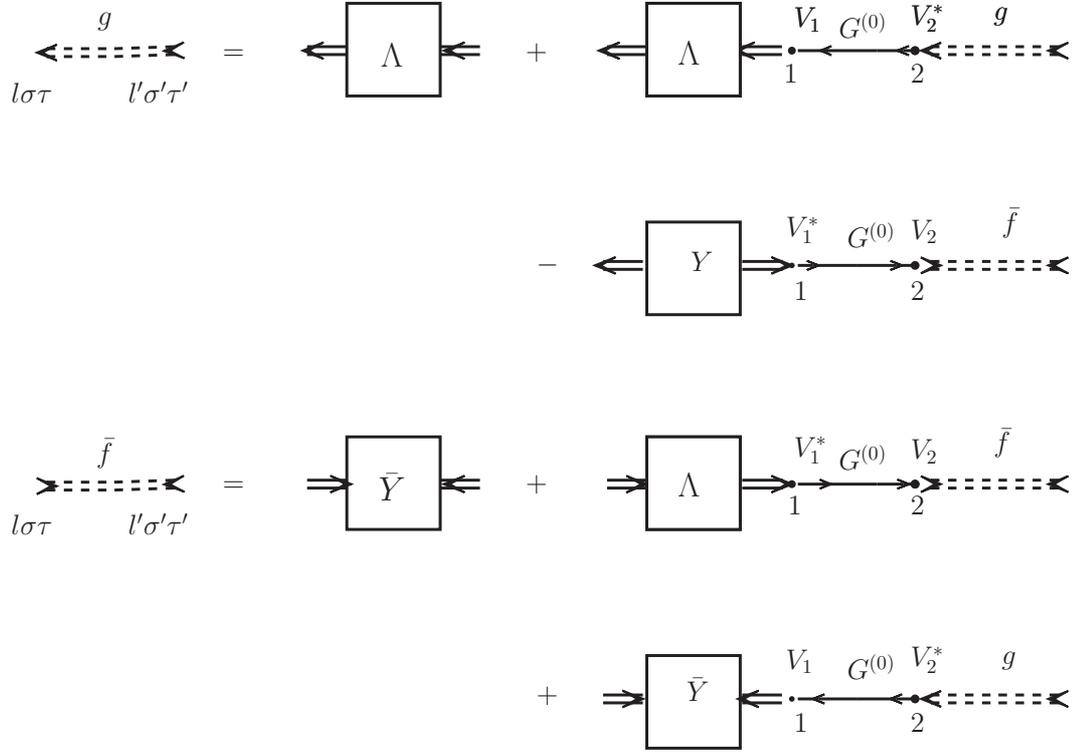}
\vspace{-0mm}
%
\caption{Dyson type equation for Green's function of impurity
electrons. $\Lambda$, $Y$, $\bar{Y}$ are correlation functions.
}\label{fig-1} \vspace{+1mm}
\end{figure*}
%

The structure representative of the diagrams in Fig. 1 is given by
the following equation
\begin{eqnarray}
\sum\limits_{\vec{k}_{1}}\sum\limits_{\vec{k}_{2}}
\frac{V_{\vec{k}_{1}l_{1}}V_{\vec{k}_{2}l_{2}}^*}{N}
G_{l_{1}\sigma_{1}l_{2}\sigma_{2}}^{0}(\vec{k}_{1}\vec{k_{2}}|\mathrm{i}\omega_{n})=
\\ \nonumber
\frac{1}{N}\sum\limits_{\vec{k}_{1}}|V_{\vec{k}_{1}l_{1}}|^{2}
G_{l_{1}\sigma_{1}}^{(0)}(\vec{k}_{1}|\mathrm{i}\omega_{n})
\delta_{l_{1}l_{2}}\delta_{\sigma_{1}\sigma_{2}}=
\delta_{l_{1}l_{2}}\delta_{\sigma_{1}\sigma_{2}}\mathcal{G}
_{l_{1}\sigma_{1}}^{(0)}(\mathrm{i}{\omega_{n}}), \label{26}
\end{eqnarray}
where
\begin{eqnarray}
\mathcal{G}_{l\sigma}^{(0)}(\mathrm{i}{\omega_{n}}) =
\frac{1}{N}\sum\limits_{\vec{k}}|V_{\vec{k}l}|^{2}
G_{l\sigma}^{(0)}(\vec{k}|\mathrm{i}\omega_{n})=
\frac{1}{N}\sum\limits_{\vec{k}}\frac{|V_{\vec{k}l}|^{2}}{\mathrm{i}\omega_{n}-\epsilon(\vec{k})}.\label{27}
\end{eqnarray}
The renormalization quantity is
\begin{eqnarray}
\mathcal{G}_{l\sigma
l^{\prime}\sigma^{\prime}}(\mathrm{i}{\omega_{n}}) =
\frac{1}{N}\sum\limits_{\vec{k}\vec{k}^{\prime}} V_{\vec{k}l}
V_{\vec{k}^{\prime}l^{\prime}}^*G_{l\sigma
l^{\prime}\sigma^{\prime}}(\vec{k}\vec{k}^{\prime}|\mathrm{i}\omega).\label{28}
\end{eqnarray}

In the Fig. 1 the double dashed lines with arrows depict
renormalized $g$ and $f$ propagators of localized electrons and
solid thin lines represent $\mathcal{G}^{0}$ function of conduction
electrons. The function $V_{1}$ means $V_{\vec{k}_{1}l_{1}}$ and
summation by repeated indices is assumed.

$\Lambda$ and $\bar{Y}$ are correlation functions. They contain a
sum of strongly connected irreducible diagrams. The simplest
examples of such diagrams are shown on Fig. 2.
%
\begin{figure*}[h]
%
\centering
\includegraphics[width=0.85\textwidth,clip]{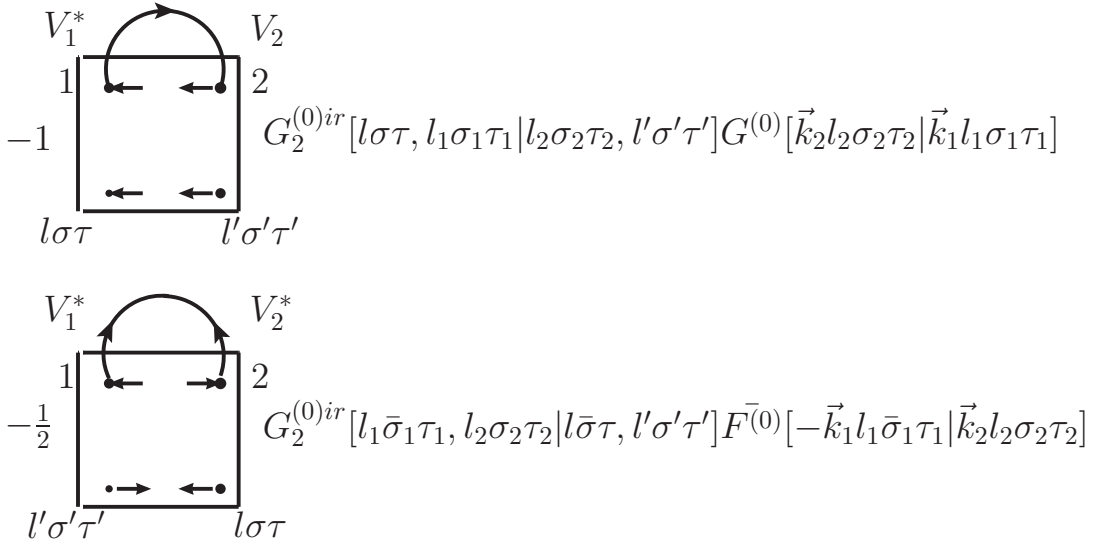}
\vspace{-0mm}
%
\caption{ The simplest examples of correlation functions $\Lambda$
and $\bar{Y}$.}\label{fig-1} \vspace{+1mm}
\end{figure*}
%

The analytical form of equations in Fig.1 is the following:
\begin{eqnarray}
g_{l\sigma l^{\prime}\sigma^{\prime}}(\mathrm{i}\omega_{n})=
\Lambda_{l\sigma l^{\prime}\sigma^{\prime}}(\mathrm{i}\omega_{n})+
\Lambda_{l\sigma l_{1}\sigma_{1}}(\mathrm{i}\omega_{n})
\mathcal{G}_{l_{1}\sigma_{1}}^{(0)}(\mathrm{i}\omega_{n})
g_{l_{1}\sigma_{1}l^{\prime}\sigma^{\prime}}(\mathrm{i}\omega_{n})-
\nonumber\\
Y_{l\sigma l_{1}\sigma_{1}}(\mathrm{i}\omega_{n})
\mathcal{G}_{l_{1}\sigma_{1}}^{(0)}(-\mathrm{i}\omega_{n})
\bar{f}_{l_{1}\sigma_{1}l^{\prime}\sigma^{\prime}}(\mathrm{i}\omega_{n}),\\
\bar{f}_{l\sigma l^{\prime}\sigma^{\prime}}(\mathrm{i}\omega_{n})=
\bar{Y}_{l\sigma l^{\prime}\sigma^{\prime}}(\mathrm{i}\omega_{n})+
\Lambda_{l_{1}\sigma_{1}l\sigma}(-\mathrm{i}\omega_{n})
\mathcal{G}_{l_{1}\sigma_{1}}^{(0)}(-\mathrm{i}\omega_{n})
\bar{f}_{l_{1}\sigma_{1}l^{\prime}\sigma^{\prime}}(\mathrm{i}\omega_{n})+
\nonumber\\ \bar{Y}_{l\sigma l_{1}\sigma_{1}}(\mathrm{i}\omega_{n})
\mathcal{G}_{l_{1}\sigma_{1}}^{(0)}(\mathrm{i}\omega_{n})
g_{l_{1}\sigma_{1}l^{\prime}\sigma^{\prime}}(\mathrm{i}\omega_{n}).
\nonumber\label{29}
\end{eqnarray}
This system of equations is rather general and admit different
phases. We shall discuss one of the most simple form with
singlet superconductivity on the paramagnetic background. \\
For this special case we use the new notations $(\bar{\sigma}=
-\sigma)$:
\begin{eqnarray}
g_{l\sigma l^{\prime}\sigma^{\prime}}(\mathrm{i}\omega_{n})&=&
\delta_{\sigma\sigma^{\prime}}g_{\sigma}^{ll^{\prime}}(\mathrm{i}\omega_{n}),\bar{f}_{l\sigma
l^{\prime}\sigma^{\prime}}(\mathrm{i}\omega_{n})=
\delta_{\sigma\bar{\sigma}^{\prime}}\bar{f}_{\bar{\sigma}\sigma}^{ll^{\prime}}(\mathrm{i}\omega_{n}),
\nonumber\\\Lambda_{l\sigma
l^{\prime}\sigma^{\prime}}(\mathrm{i}\omega_{n})&=&
\delta_{\sigma\sigma^{\prime}}\Lambda_{\sigma}^{ll^{\prime}}(\mathrm{i}\omega_{n}),
\bar{Y}_{l\sigma l^{\prime}\sigma^{\prime}}(\mathrm{i}\omega_{n})=
\delta_{\sigma\bar{\sigma}^{\prime}}\bar{Y}_{\bar{\sigma}\sigma}^{ll^{\prime}}(\mathrm{i}\omega_{n}),
\\g_{l\sigma}^{(0)}(\mathrm{i}\omega_{n})&=&g_{\sigma}^{(0)l}(\mathrm{i}\omega_{n}).
\nonumber\label{30}
\end{eqnarray}

By using these definitions we obtain:
\begin{eqnarray}
g_{\sigma}^{ll^{\prime}}(\mathrm{i}\omega_{n})&=&
\Lambda_{\sigma}^{ll^{\prime}}(\mathrm{i}\omega_{n})+
\Lambda_{\sigma}^{ll_{1}}(\mathrm{i}\omega_{n})
\mathcal{G}_{\sigma}^{l_{1}(0)}(\mathrm{i}\omega_{n})g_{\sigma}^{l_{1}l^{\prime}}(\mathrm{i}\omega_{n})-
Y_{\sigma\bar{\sigma}}^{ll_{1}}(\mathrm{i}{\omega_{n}})
\mathcal{G}_{\bar{\sigma}}^{l_{1}(0)}(-\mathrm{i}\omega_{n})
\bar{f}_{\bar{\sigma}\sigma}^{l_{1}l^{\prime}}(\mathrm{i}\omega_{n}),
\\ \nonumber\bar{f}_{\bar{\sigma}\sigma}^{ll^{\prime}}(\mathrm{i}\omega_{n})&=&
\bar{Y}_{\bar{\sigma}\sigma}^{ll^{\prime}}(\mathrm{i}\omega_{n})+
\Lambda_{\bar{\sigma}}^{l_{1}l}(-\mathrm{i}\omega_{n})
\mathcal{G}_{\bar{\sigma}}^{l_{1}(0)}(-\mathrm{i}\omega_{n})
\bar{f}_{\bar{\sigma}\sigma}^{l_{1}l^{\prime}}(\mathrm{i}\omega_{n})
+ \bar{Y}_{\bar{\sigma}\sigma}^{ll_{1}}(\mathrm{i}\omega_{n})
\mathcal{G}_{\sigma}^{{l_{1}}(0)}(\mathrm{i}\omega_{n})
g_{\sigma}^{l_{1}l^{\prime}}(\mathrm{i}\omega_{n}). \label{31}
\end{eqnarray}
In the absence of orbital degeneracy this system of equation has the
known solution [14]
\begin{eqnarray}
g_{\sigma}(\mathrm{i}\omega_{n})&=&
\frac{\Lambda_{\sigma}(\mathrm{i}\omega_{n})-\mathcal{G}_{\bar{\sigma}}^{(0)}(-\mathrm{i}\omega_{n})
[\Lambda_{\sigma}(\mathrm{i}\omega_{n})\Lambda_{\bar{\sigma}}(-\mathrm{i}\omega_{n})+
Y_{\sigma\bar{\sigma}}(\mathrm{i}\omega_{n})\bar{Y}_{\bar{\sigma}\sigma}(\mathrm{i}\omega_{n})]}
{d_{\sigma}(\mathrm{i}\omega_{n})}, \nonumber\\
\bar{f}_{\bar{\sigma}\sigma}(\mathrm{i}\omega_{n})&=&
\frac{\bar{Y}_{\bar{\sigma}\sigma}(\mathrm{i}\omega_{n})}{d_{\sigma}(\mathrm{i}\omega_{n})},
\,\ \,\ f_{\sigma\bar{\sigma}}(\mathrm{i}\omega_{n})=
\frac{Y_{\sigma\bar{\sigma}}(\mathrm{i}\omega_{n})}{d_{\sigma}(\mathrm{i}\omega_{n})},
\\ \nonumber d_{\sigma}(\mathrm{i}\omega_{n})&=&
(1-\Lambda_{\sigma}(\mathrm{i}\omega_{n})\mathcal{G}_{\sigma}^{(0)}(\mathrm{i}\omega_{n}))
(1-\Lambda_{\bar{\sigma}}(-\mathrm{i}\omega_{n})\mathcal{G}_{\bar{\sigma}}^{(0)}(-\mathrm{i}\omega_{n}))
+\\
&&\mathcal{G}_{\sigma}^{(0)}(\mathrm{i}\omega_{n})\mathcal{G}_{\bar{\sigma}}^{(0)}(-\mathrm{i}\omega_{n})
Y_{\sigma\bar{\sigma}}(\mathrm{i}\omega_{n})\bar{Y}_{\bar{\sigma}\sigma}(\mathrm{i}\omega_{n}).\nonumber
\label{32}
\end{eqnarray}
Solutions of the equation $(31)$ for the normal state of the
degenerate system has the form:
\begin{eqnarray}
g_{\sigma}^{11}(\mathrm{i}\omega_{n})&=&
\frac{\Lambda_{\sigma}^{11}(\mathrm{i}\omega_{n})-
\mathcal{G}_{\sigma}^{2(0)}(\mathrm{i}\omega_{n})
[\Lambda_{\sigma}^{11}(\mathrm{i}\omega_{n})\Lambda_{\sigma}^{22}(\mathrm{i}\omega_{n})
-\Lambda_{\sigma}^{12}(\mathrm{i}\omega_{n})\Lambda_{\sigma}^{21}(\mathrm{i}\omega_{n})]}
{d_{\sigma}(\mathrm{i}\omega_{n})}, \nonumber\\
g_{\sigma}^{21}(\mathrm{i}\omega_{n})&=&
\frac{\Lambda_{\sigma}^{21}(\mathrm{i}\omega_{n})}{d_{\sigma}(\mathrm{i}\omega_{n})},\quad
d_{\sigma}(\mathrm{i}\omega_{n})=
(1-\mathcal{G}_{\sigma}^{1(0)}(\mathrm{i}\omega_{n})\times\\
&&\Lambda_{\sigma}^{11}(\mathrm{i}\omega_{n}))
(1-\mathcal{G}_{\sigma}^{2(0)}(\mathrm{i}\omega_{n})\Lambda_{\sigma}^{22}(\mathrm{i}\omega_{n}))-
\mathcal{G}_{\sigma}^{1(0)}(\mathrm{i}\omega_{n})\mathcal{G}_{\sigma}^{2(0)}(\mathrm{i}\omega_{n})
\Lambda_{\sigma}^{12}(\mathrm{i}\omega_{n})\Lambda_{\sigma}^{21}(\mathrm{i}\omega_{n}).
\nonumber \label{33}
\end{eqnarray}
The other two functions are obtained by changing the indexes
$1\leftrightarrow 2$. These equations are of Dyson type. They
determine Green's functions through correlation functions
$\Lambda=g^{(0)}+Z$, $Y$ and $\bar{Y}$ ones. The last three can only
be given in a form of infinite diagram series, since the exact
solution does not exist.

An example of efficient summation of diagram and determination of
the correlation function $Z$, $Y$ and $\bar{Y}$ is presented on the
Fig. 3.
%
\begin{figure*}[h]
%
\centering
\includegraphics[width=0.85\textwidth,clip]{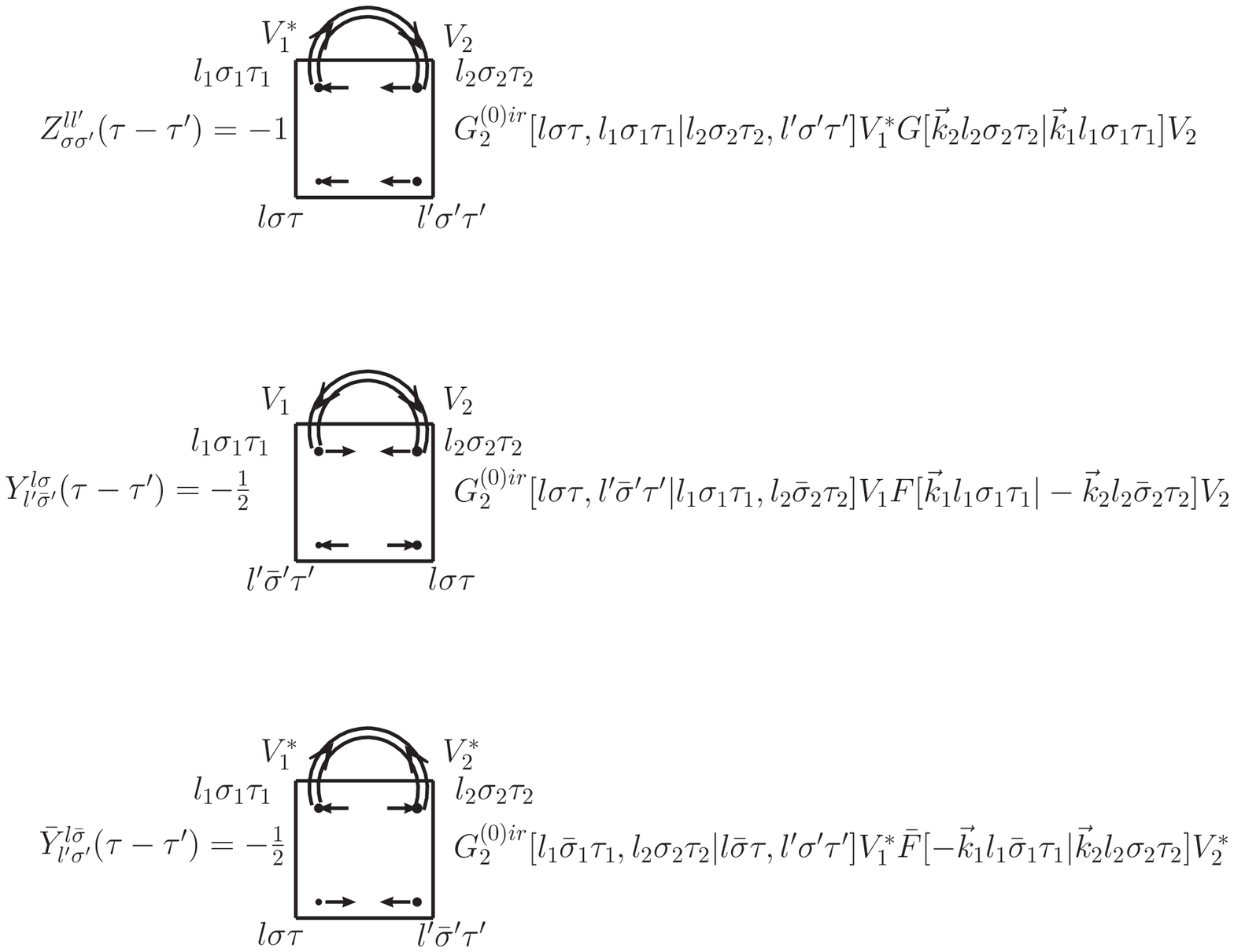}
\vspace{-0mm}
%
\caption{ The main approximation for the correlation functions. The
solid double lines with arrows depict the full Green's functions of
conduction electrons. The rectangles depict the irreducible Green's
functions of the impurity electrons.}\label{fig-1} \vspace{+1mm}
\end{figure*}
%

The diagrams of Fig. 3 differ from the ones of Fig. 2 by the
presence of the full conduction electron Green's function instead of
the bare one of Fig. 2. This difference is the result of ladder
summation of main diagrams.
\section{Correlation functions}

The simplest correlation function is determined as
\begin{eqnarray}
G_{2}^{irr}[1,2|\bar{3},\bar{4}]=g_{2}^{(0)}(1,2|\bar{3},\bar{4})-g_{1}^{(0)}(1|\bar{4})g_{1}^{(0)}(2|\bar{3})
+g_{1}^{(0)}(1|\bar{3})g_{1}^{(0)}(2|\bar{4}), \\
g_{2}^{(0)}(1,2|\overline{3},\overline{4})= \langle T
d_{1}d_{2}\bar{d}_{3}\bar{d}_{4}\rangle_{0},g_{1}^{(0)}(1|\bar{4})=
-\langle Td_{1}\bar{d}_{4}\rangle_{0},
1=(l_{1},\sigma_{1},\tau_{1}),\nonumber\label{34}
\end{eqnarray}
with two- and one-particle bare Green's functions of localized
electrons.

Because the presence of the Coulomb interactions in zero order
Hamiltonian, equation (34) is different of zero and contains charge,
spin and pairing fluctuations.

The two-particle Green's function $g_{2}^{(0)}$ is the sum of 4!
terms of different time ordered electron operators products. The
statistical averages of these quantities are calculated by using
Hubbard transfer operators representation.

We need the Fourier representation of these functions
\begin{eqnarray}
G_{2}^{irr}[l_{1}\sigma_{1}\tau_{1};l_{2}\sigma_{2}\tau_{2}|l_{3}\sigma_{3}\tau_{3};l_{4}\sigma_{4}\tau_{4}]&=&\frac{1}{\beta^{4}}
\sum\limits_{\omega_{1}\omega_{2}\omega_{3}\omega_{4}}G_{2}^{irr}[l_{1}\sigma_{1}\mathrm{i}\omega_{1};l_{2}\sigma_{2}\mathrm{i}
\omega_{2}|l_{3}\sigma_{3}\mathrm{i}\omega_{3};
l_{4}\sigma_{4}\mathrm{i}\omega_{4}]\times\nonumber
\\e^{-\mathrm{i}\omega_{1}\tau_{1}-\mathrm{i}\omega_{2}\tau_{2}+\mathrm{i}\omega_{3}\tau_{3}+\mathrm{i}\omega_{4}\tau_{4}}
,\nonumber\\g_{1}^{(0)}(l_{1}\sigma_{1}\tau_{1}|l_{2}\sigma_{2}\tau_{2})&=&
\frac{1}{\beta}\sum\limits_{\omega_{1}}g_{1}^{(0)}(l_{1}\sigma_{1};l_{2}\sigma_{2}|\mathrm{i}\omega_{1})e^{-\mathrm{i}\omega_{1}(\tau_{1}-\tau_{2})}\\
g_{1}^{(0)}(l_{1}\sigma_{1};l_{2}\sigma_{2}|\mathrm{i}\omega_{1})&\approx&\delta_{l_{1}l_{2}}\delta_{\sigma_{1}\sigma_{2}}m(\mathrm{i}\omega_{1})=
\frac{\delta_{l_{1}l_{2}}\delta_{\sigma_{1}\sigma_{2}}}{2}\left(\frac{1}{\mathrm{i}\omega_{1}+E_{2}-E_{9}}+\frac{1}{\mathrm{i}\omega_{1}+E_{9}-E_{12}}\right).
\nonumber\label{35}
\end{eqnarray}
\begin{eqnarray}
G_{2}^{irr}[l_{1}\sigma_{1}\mathrm{i}\omega_{1};l_{2}\sigma_{2}\mathrm{i}\omega_{2}|l_{3}\sigma_{3}\mathrm{i}\omega_{3};l_{4}\sigma_{4}\mathrm{i}\omega_{4}]
=g_{2}^{(0)}[l_{1}\sigma_{1}\mathrm{i}\omega_{1};l_{2}\sigma_{2}\mathrm{i}\omega_{2}|l_{3}\sigma_{3}\mathrm{i}
\omega_{3};l_{4}\sigma_{4}\mathrm{i}\omega_{4}]\nonumber-\\
\beta\delta(\omega_{1}+\omega_{2}-\omega_{3}-\omega_{4})[\beta\delta(\omega_{1}-\omega_{4})g_{1}^{(0)}(l_{1}
\sigma_{1};l_{4}\sigma_{4}|\mathrm{i}\omega_{1})g_{1}^{(0)}(l_{2}
\sigma_{2};l_{3}\sigma_{3}|\mathrm{i}\omega_{2})-\\\beta\delta(\omega_{1}-\omega_{3})g_{1}^{(0)}(l_{1}
\sigma_{1};l_{3}\sigma_{3}|\mathrm{i}\omega_{1})g_{1}^{(0)}(l_{2}
\sigma_{2};l_{4}\sigma_{4}|\mathrm{i}\omega_{2})].
\nonumber\label{36}
\end{eqnarray}

There exists the law of frequency conservation
\begin{eqnarray}
G_{2}^{irr}[l_{1}\sigma_{1}\mathrm{i}\omega_{1};l_{2}\sigma_{2}\mathrm{i}\omega_{2}|l_{3}
\sigma_{3}\mathrm{i}\omega_{3};l_{4}\sigma_{4}\mathrm{i}\omega_{4}]
=\\
\beta\delta(\omega_{1}+\omega_{2}-\omega_{3}-\omega_{4})\widetilde{G}_{2}^{irr}[l_{1}\sigma_{1}
\mathrm{i}\omega_{1};l_{2}\sigma_{2}\mathrm{i}\omega_{2}|l_{3}\sigma_{3}\mathrm{i}\omega_{3};l_{4}\sigma_{4}\mathrm{i}\omega_{4}].
\nonumber\label{37}
\end{eqnarray}

The statistical averages of chronologically ordered products of the electron operators of the function $g_{2}^{(0)}$
have different weights of the form $\frac{e^{-\beta E_{n}}}{Z_{0}}$, where $E_{n}$ are the energies determined in previous section.
Because $E_{9}$ is the lowest energy the weight $e^{-\beta E_{9}}$ is the main of them and only such terms are taken into account.

Just such considerations determined us to use instead initial exact
equation (16) for zero order Green's function $g^{(0)}_{l\sigma
l^{\prime}\sigma^{\prime}}$ the approximate value (35). Zero order
partition function $Z_{0}$ (17) concomitant is approximated as
$3e^{-\beta E_{9}}$.

For example the contribution to function
$g_{2}^{(0)}[l_{1}\sigma_{1}\mathrm{i}\omega_{1};l_{2}\sigma_{2}\mathrm{i}\omega_{2}|l_{3}
\sigma_{3}\mathrm{i}\omega_{3};l_{4}\sigma_{4}\mathrm{i}\omega_{4}]$
with time order $\beta>\tau_{1}>\tau_{3}>\tau_{2}>\tau_{4}>0$ and
with weight $e^{-\beta E_{9}}$ is
\begin{eqnarray}
-\delta_{l_{1}l_{3}}\delta_{l_{2}l_{4}}(\frac{1}{4}\delta_{\sigma_{1}\sigma_{3}}\delta_{\sigma_{2}\sigma_{4}}+
\delta_{\sigma_{1},-\sigma_{3}}\delta_{\sigma_{2},-\sigma_{4}}\delta_{\sigma_{2}\sigma_{3}}
+\delta_{\sigma_{1}\sigma_{3}}\delta_{\sigma_{2}\sigma_{4}}
\delta_{\sigma_{1}\sigma_{4}}\delta_{\sigma_{2}\sigma_{3}}
)I_{1\bar{3}2\bar{4}}^{(1)}\nonumber-\\( \delta_{3-l_{1}-l_{3},0}\delta_{3-l_{2}-l_{4},0}+(-1)^{l_{1}+l_{4}}
\delta_{l_{1}l_{3}}\delta_{l_{2}l_{4}})(\frac{1}{4}\sigma_{1}\sigma_{4}
\delta_{\sigma_{1}\sigma_{3}}\delta_{\sigma_{2}\sigma_{4}}+\frac{1}{2}
\delta_{\sigma_{1},-\sigma_{3}}\delta_{\sigma_{2},-\sigma_{4}}\delta_{\sigma_{1}\sigma_{4}})I_{1\bar{3}2\bar{4}}^{(2)}-\\
(-1)^{l_{1}+l_{4}}\delta_{3-l_{1}-l_{3},0}\delta_{3-l_{2}-l_{4},0}(\frac{1}{4}\sigma_{1}\sigma_{4}
\delta_{\sigma_{1}\sigma_{3}}\delta_{\sigma_{2}\sigma_{4}}+\frac{1}{2}
\delta_{\sigma_{1},-\sigma_{3}}\delta_{\sigma_{2},-\sigma_{4}}\delta_{\sigma_{1}\sigma_{4}})I_{1\bar{3}2\bar{4}}^{(3)},
\nonumber\label{38}
\end{eqnarray}
where
\begin{eqnarray}
I_{1\bar{3}2\bar{4}}^{(1)}=\frac{e^{-\beta
E_{9}}}{Z_{0}}\int\limits_{0}^{\beta}d\tau_{1}\int\limits_{0}^{\tau_{1}}d\tau_{3}\int\limits_{0}^{\tau_{3}}d\tau_{2}
\int\limits_{0}^{\tau_{2}}d\tau_{4}e^{(E_{9}-E_{12})(\tau_{1}+\tau_{2}-\tau_{3}-\tau_{4})}e^{\mathrm{i}\omega_{1}\tau_{1}
+\mathrm{i}\omega_{2}\tau_{2}-\mathrm{i}\omega_{3}\tau_{3}-\mathrm{i}\omega_{4}\tau_{4}},\nonumber\\
I_{1\bar{3}2\bar{4}}^{(2)}=\frac{e^{-\beta
E_{9}}}{Z_{0}}\int\limits_{0}^{\beta}d\tau_{1}\int\limits_{0}^{\tau_{1}}d\tau_{3}\int\limits_{0}^{\tau_{3}}d\tau_{2}
\int\limits_{0}^{\tau_{2}}d\tau_{4}e^{(E_{9}-E_{12})(\tau_{1}-\tau_{4})+(E_{6}-E_{12})(\tau_{2}-\tau_{3})}
e^{\mathrm{i}\omega_{1}\tau_{1}+\mathrm{i}\omega_{2}\tau_{2}-\mathrm{i}\omega_{3}\tau_{3}-\mathrm{i}\omega_{4}\tau_{4}},\\
I_{1\bar{3}2\bar{4}}^{(3)}=\frac{e^{-\beta
E_{9}}}{Z_{0}}\int\limits_{0}^{\beta}d\tau_{1}\int\limits_{0}^{\tau_{1}}d\tau_{3}\int\limits_{0}^{\tau_{3}}d\tau_{2}
\int\limits_{0}^{\tau_{2}}d\tau_{4}e^{(E_{9}-E_{12})(\tau_{1}-\tau_{4})+(E_{7}-E_{12})(\tau_{2}-\tau_{3})}e^{\mathrm{i}\omega_{1}\tau_{1}
+\mathrm{i}\omega_{2}\tau_{2}-\mathrm{i}\omega_{3}\tau_{3}-\mathrm{i}\omega_{4}\tau_{4}}.
\nonumber\label{39}
\end{eqnarray}

These 4-fold multiple integrals by time variable $\tau$ can be
transformed in contour integral by using the method of Claude Bloch
[15]. With this purpose it is necessary to introduce the exponential
form
\begin{eqnarray}
e^{(\beta-\tau_{1})\bar{E}_{0}+(\tau_{1}-\tau_{3})\bar{E}_{1}+(\tau_{3}-\tau_{2})\bar{E}_{2}+(\tau_{2}-\tau_{4})\bar{E}_{3}
+(\tau_{4}-0)\bar{E}_{4}},\label{40}
\end{eqnarray}
which must be compared with exponential form of our integrals
$I_{1\bar{3}2\bar{4}}^{(n)}$. Comparison with
$I_{1\bar{3}2\bar{4}}^{(1)}$ give us the result
\begin{eqnarray}
\bar{E}_{0}=-{E}_{9},\bar{E}_{2}=-{E}_{9}+\mathrm{i}\omega_{1}-\mathrm{i}\omega_{3},\bar{E}_{4}=-{E}_{9}+\mathrm{i}\Omega,
\bar{E}_{1}=-{E}_{12}+\mathrm{i}\omega_{1},\\\bar{E}_{3}=-{E}_{12}+\mathrm{i}\omega_{1}+\mathrm{i}\omega_{2}-\mathrm{i}\omega_{3},\Omega=
\omega_{1}+\omega_{2}-\omega_{3}-\omega_{4}.\nonumber\label{41}
\end{eqnarray}

Our integral $I_{1\bar{3}2\bar{4}}^{(1)}$ is transformed in the contour integral
\begin{eqnarray}
I^{(1)}=\frac{1}{2\pi\mathrm{i}}\frac{1}{Z_{0}}\oint\limits_{C^{+}}\frac{dz
e^{-\beta z}}{(z+\bar{E}_{0})
(z+\bar{E}_{1})(z+\bar{E}_{2})(z+\bar{E}_{3})(z+\bar{E}_{4})},\label{42}
\end{eqnarray}
where contour $C^{+}$ surrounds the real axis in the positive
direction. The integrals $I^{(2)}$ and $I^{(3)}$ have the same form
(42) but differ in the definition of energy $\bar{E}_{2}$. For
$I^{(2)}$ the energy
$\bar{E}_{2}=-{E}_{6}+\mathrm{i}\omega_{1}-\mathrm{i}\omega_{3}$ and
for $I^{(3)}$,
$\bar{E}_{2}=-{E}_{7}+\mathrm{i}\omega_{1}-\mathrm{i}\omega_{3}$.
Other parameters coincide.

The contour integral (42) is evaluated by the method of residues. The simple results are obtained when the parameters
$\bar{E}_{n}$ are different. The existence of multiple poles is possible for the special values of frequencies $\omega_{n}.$

For example in the case when $\omega_{1}-\omega_{3}=0$ and $\Omega=0$ we have $\bar{E}_{0}=\bar{E}_{2}=\bar{E}_{4}$ and
the pole $z=-\bar{E}_{0}$ is 3-fold multiple with the residue
\begin{eqnarray}
\frac{1}{2}\left(\frac{e^{-\beta
Z}}{(z+\bar{E}_{1})(z+\bar{E}_{3})}\right)^{\prime\prime}_{z=-\bar{E}_{0}}.\label{43}
\end{eqnarray}

To find all possible multiple poles we consider different values of frequencies using the identity $1=\delta(\omega)+\psi(\omega)$, where $\psi(\omega)=1-\delta(\omega)$.
For example we consider the possibility when $\Omega$ can be equal to zero and $\omega_{1}=\omega_{3}$. We have the identity:
\begin{eqnarray}
1=(\delta(\Omega)+\psi(\Omega))(\delta(\omega_{1}-\omega_{3})+\psi(\omega_{1}-\omega_{3}))=\\
\delta(\Omega)\delta(\omega_{1}-\omega_{3})+\delta(\Omega)\psi(\omega_{1}-\omega_{3}))+
\psi(\Omega)\delta(\omega_{1}-\omega_{3})+\psi(\Omega)\psi(\omega_{1}-\omega_{3}))\nonumber
.\label{44}
\end{eqnarray}
The first term in the right-hand part of this equation admits the
existence of triple pole, the next two terms admit double poles and
last term admit double and single poles.

We shall take into account these residues, statistical weights of
which is $\frac{e^{-\beta E_{9}}}{Z_{0}}$, and shall omit the other
ones. In such approximation we have
\begin{eqnarray}
Z_{0}I_{1\bar{3}2\bar{4}}^{(1)}=\frac{1}{2}\delta(\Omega)\delta(\omega_{1}-\omega_{3})\left(\frac{e^{-\beta
Z}}{(z+\bar{E}_{1})(z+\bar{E}_{3})}
\right)^{\prime\prime}_{z=-\bar{E}_{0}}+\nonumber\\
\delta(\Omega)\psi(\omega_{1}-\omega_{3})\left
[\left(\frac{e^{-\beta Z}}{(z+\bar{E}_{1})(z+\bar{E}_{2})
(z+\bar{E}_{3})}\right)^{\prime}_{z=-\bar{E}_{0}}+\left(\frac{e^{-\beta Z}}
{(z+\bar{E}_{0})^{2}(z+\bar{E}_{1})(z+\bar{E}_{3})}\right)_{z=-\bar{E}_{2}}\right]+\nonumber\\
\delta(\omega_{1}-\omega_{3})\psi(\Omega)\left[\left(\frac{e^{-\beta
Z}}{(z+\bar{E}_{1})
(z+\bar{E}_{3})(z+\bar{E}_{4})}\right)^{\prime}_{z=-\bar{E}_{0}}+\left(\frac{e^{-\beta Z}}{(z+\bar{E}_{0})^{2}(z+\bar{E}_{1})(z+\bar{E}_{3})}
\right)_{z=-\bar{E}_{4}}\right]+\nonumber\\
\psi(\Omega)\psi(\omega_{1}-\omega_{3})\delta(\omega_{2}-\omega_{4})\times\\
\left [\left(\frac{e^{-\beta Z}}{(z+\bar{E}_{0})(z+\bar{E}_{1})
(z+\bar{E}_{3})}\right)^{\prime}_{z=-\bar{E}_{2}}+\left(\frac{e^{-\beta
Z}}{(z+\bar{E}_{2})^{2}(z+\bar{E}_{1})(z+\bar{E}_{3})}\right)_{z=-\bar{E}_{0}}\right]+\nonumber\\
\psi(\Omega)\psi(\omega_{1}-\omega_{3})\psi(\omega_{2}-\omega_{4})
[\left(\frac{e^{-\beta Z}}{(z+\bar{E}_{1})(z+\bar{E}_{2})(z+\bar{E}_{3})(z+\bar{E}_{4})}\right)_{z=-\bar{E}_{0}}+\nonumber\\
\left(\frac{e^{-\beta
Z}}{(z+\bar{E}_{0})(z+\bar{E}_{1})(z+\bar{E}_{3})(z+\bar{E}_{4})}\right)_{z=-\bar{E}_{2}}+
\left(\frac{e^{-\beta
Z}}{(z+\bar{E}_{0})(z+\bar{E}_{1})(z+\bar{E}_{2})(z+\bar{E}_{3})}\right)_{z=-\bar{E}_{4}}]\nonumber
.\label{45}
\end{eqnarray}

The contribution of other poles is negligible. Our next
approximation consists in preserving, in the case of low
temperature, of the main part of the second derivative (43) just of
the form
\begin{eqnarray}
\Delta I=\frac{\beta^{2}e^{-\beta
E_{9}}}{2Z_{0}(\bar{E}_{1}-\bar{E}_{0})(\bar{E}_{3}-\bar{E}_{0})}.\label{46}
\end{eqnarray}

This contribution together with contribution (36) of the product of
one-particle Green's functions determines the main part of the
correlation function. This part is designed as $G_{2}^{(0)irr}$.

After some transformation and summation of different contributions
we obtain the main approximation for the correlation function:
\begin{eqnarray}
G_{2}^{(0)irr}[l_{1}\sigma_{1}\mathrm{i}\omega_{1};l_{2}\sigma_{2}\mathrm{i}\omega_{2}|l_{3}\sigma_{3}\mathrm{i}\omega_{3};l_{4}\sigma_{4}\mathrm{i}\omega_{4}]=\frac{\beta}{6}
\delta(\omega_{1}+\omega_{2}-\omega_{3}-\omega_{4})p(\mathrm{i}\omega_{1})p(\mathrm{i}\omega_{2})\times\nonumber\\
({\beta}\delta(\omega_{1}-\omega_{4})\delta_{l_{1}l_{4}}\delta_{l_{2}l_{3}}[2\delta_{\sigma_{1},
-\sigma_{4}}\delta_{\sigma_{2},-\sigma_{3}}\delta_{\sigma_{2}\sigma_{4}}+
\delta_{\sigma_{1}\sigma_{3}}\delta_{\sigma_{1}\sigma_{4}}
\delta_{\sigma_{2}\sigma_{3}}-\delta_{\sigma_{1}\sigma_{4}}\delta_{\sigma_{2}\sigma_{3}}
\delta_{\sigma_{3},-\sigma_{1}}]-\\{\beta}\delta(\omega_{1}-\omega_{3})\delta_{l_{1}l_{3}}
\delta_{l_{2}l_{4}}[2\delta_{\sigma_{1},-\sigma_{3}}\delta_{\sigma_{2},-\sigma_{4}}\delta_{\sigma_{2}\sigma_{3}}+
\delta_{\sigma_{1}\sigma_{3}}\delta_{\sigma_{2}\sigma_{4}}
\delta_{\sigma_{1}\sigma_{4}}-\delta_{\sigma_{1}\sigma_{3}}\delta_{\sigma_{2}\sigma_{4}}
\delta_{\sigma_{4},-\sigma_{1}}]), \nonumber\label{47}
\end{eqnarray}
with
\begin{eqnarray}
p(\mathrm{i}\omega)=
\left(\frac{1}{\mathrm{i}\omega+E_{2}-E_{9}}-\frac{1}{\mathrm{i}\omega+E_{9}-E_{12}}\right).\label{48}
\end{eqnarray}

\section{Mott-Hubbard phase transition}

As has been mentioned above, one example of efficient summation of
diagrams which determine correlation function $Z$ and $\Lambda$ is
presented on the Fig. 4. It has the form
%
\begin{figure*}[h]
%
\centering
\includegraphics[width=0.85\textwidth,clip]{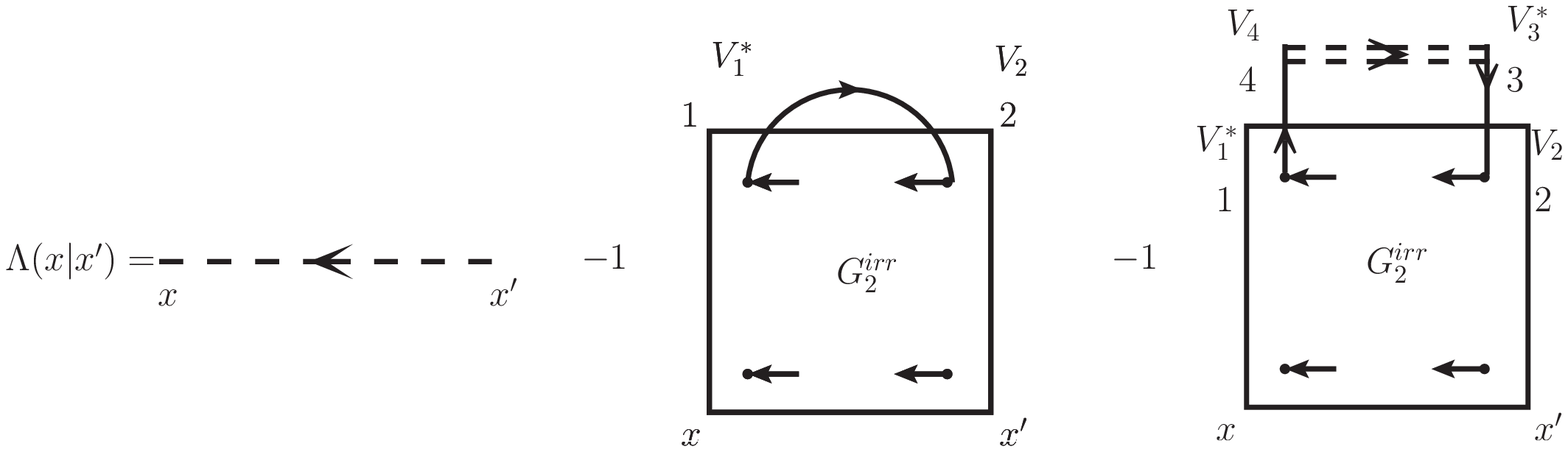}
\vspace{-0mm}
%
\caption{ The main equation for the function
$\Lambda(x|x^{\prime}).$ Here $x$ is $(l,\sigma,\mathrm{i}\omega)$.
The thin dashed line represents the bare local one-particle Green's
function and the double dashed the renormalized one. The thin solid
line represents the conduction propagator.}\label{fig-4}
\vspace{+1mm}
\end{figure*}
%

First of all we shall discuss the approximation with zero order
correlation function $G_{2}^{(0)irr}.$ Using the result (47) we
obtain
\begin{eqnarray}
\frac{1}{\beta}\sum_{\omega_{1}}\sum_{l_{1}\sigma_{1}}\widetilde{G}_{2}^{(0)irr}
[l\sigma \mathrm{i} \omega;l_{1}\sigma_{1}\mathrm{i}
\omega_{1}|l_{1}\sigma_{1}\mathrm{i}
\omega_{1};l^{\prime}\sigma^{\prime}
\mathrm{i}\omega]\mathcal{G}_{l_{1}\sigma_{1}}^{(0)}(\mathrm{i}\omega_{1})=
-\frac{1}{2}\delta_{\sigma\sigma^{\prime}}\delta_{l
l^{\prime}}[p(\mathrm{i}\omega)]^2\mathcal{G}_{l\sigma}^{(0)}(\mathrm{i}\omega),\label{49}
\end{eqnarray}
\begin{eqnarray}
\sum_{l_{1}\sigma_{1}l_{2}\sigma_{2}}\widetilde{G}_{2}^{(0)irr} [l\sigma
\mathrm{i} \omega;l_{1}\sigma_{1}\mathrm{i}
\omega_{1}|l_{2}\sigma_{2}\mathrm{i}
\omega_{1};l^{\prime}\sigma^{\prime}\mathrm{i}\omega]
\mathcal{G}_{l_{2}\sigma_{2}}^{(0)}(\mathrm{i}\omega_{1})
g_{l_{2}\sigma_{2}l_{1}\sigma_{1}}(\mathrm{i}\omega_{1})
\mathcal{G}_{l_{1}\sigma_{1}}^{(0)}(\mathrm{i}\omega_{1})=\nonumber \\
\frac{1}{6}p(\mathrm{i}\omega)p(\mathrm{i}\omega_{1})\{
\beta\delta(\omega-\omega)\delta_{ll^{\prime}} \sum_{l_{1}}
(2\delta_{\sigma^{\prime},-\sigma}g_{l_{1}\sigma}^{(0)}(\mathrm{i} \omega_{1}) \nonumber \\
g_{l_{1}\sigma l_{1},-\sigma}(\mathrm{i}\omega_{1})
\mathcal{G}_{l_{1},-\sigma}^{(0)}(\mathrm{i}\omega_{1}) +
\delta_{\sigma\sigma^{\prime}}
\mathcal{G}_{l_{1}\sigma}^{(0)}(\mathrm{i}\omega_{1}) g_{l_{1}\sigma
l_{1}\sigma}(\mathrm{i}\omega_{1})
\mathcal{G}_{l_{1}\sigma}^{(0)}(\mathrm{i}\omega_{1}) - \nonumber \\
-\delta_{\sigma\sigma^{\prime}}
\mathcal{G}_{l_{1},-\sigma}^{(0)}(\mathrm{i}\omega_{1})g_{l_{1},-\sigma
l_{1,}-\sigma}(\mathrm{i}\omega_{1})\mathcal{G}_{l_{1},-\sigma}^{(0)}(\mathrm{i}\omega_{1})
) \\-\beta \delta(\omega - \omega_{1}) [2
\delta_{\sigma\sigma^{\prime}}\mathcal{G}_{l,-\sigma}^{(0)}(\mathrm{i}\omega_{1})
g_{l,-\sigma l^{\prime},-\sigma}(\mathrm{i}\omega_{1})\mathcal{G}_{l^{\prime},-\sigma}^{(0)}(\mathrm{i}\omega_{1})  + \nonumber \\
+\delta_{\sigma\sigma^{\prime}}\mathcal{G}_{l\sigma}^{(0)}(\mathrm{i}\omega_{1})
g_{l\sigma
l^{\prime}\sigma}(\mathrm{i}\omega_{1})\mathcal{G}_{l^{\prime}\sigma}^{(0)}
(\mathrm{i}\omega_{1})-\delta_{\sigma-\sigma^{\prime}}\mathcal{G}_{l\sigma}^{(0)}(\mathrm{i}\omega_{1})
g_{l\sigma
l^{\prime}\sigma^{\prime}}(\mathrm{i}\omega_{1})\mathcal{G}_{l^{\prime}\sigma^{\prime}}^{(0)}(\mathrm{i}\omega_{1})
] \}.\nonumber\label{50}
\end{eqnarray}%
We keep the terms which preserve the spin and have the form
$\delta_{\sigma \sigma_{\prime}}$ and omit the terms with spin-flipp
of the form $\delta_{\sigma^{\prime},-\sigma}$ and also omit the
terms which are reciprocally subtracted and differ only by the sign
of spin. We take into account that the function
$\mathcal{G}_{l}^{(0)}(\mathrm{i}\omega)$ doesn't depend of spin
index and
\begin{equation}
\sum_{\sigma_{1}}\sigma_{1}\sigma
\mathcal{G}_{l\sigma_{1}}^{(0)}(\mathrm{i}\omega)=0.\label{51}
\end{equation}
As a result of such simplifications we obtain
\begin{eqnarray}
\sum_{l_{1}\sigma_{1}l_{2}\sigma_{2}}\widetilde{G}_{2}^{(0)irr} [l\sigma
\mathrm{i} \omega;l_{1}\sigma_{1}\mathrm{i}
\omega_{1}|l_{2}\sigma_{2}\mathrm{i}
\omega_{2};l^{\prime}\sigma^{\prime}\mathrm{i}\omega]
\mathcal{G}_{l_{2}\sigma_{2}}^{(0)}(\mathrm{i}\omega_{1})
g_{l_{2}\sigma_{2}l_{1}\sigma_{1}}(\mathrm{i}\omega_{1})
\mathcal{G}_{l_{1}\sigma_{1}}^{(0)}(\mathrm{i}\omega_{1})=\\
-\frac{1}{2}\beta \delta(\omega -
\omega_{1})p(\mathrm{i}\omega)p(\mathrm{i}\omega_{1})\delta_{\sigma\sigma^{\prime}}
\mathcal{G}_{l\sigma}^{(0)}(\mathrm{i}\omega_{1}) g_{l\sigma
l^{\prime}\sigma}(\mathrm{i}\omega_{1})\mathcal{G}_{l^{\prime}\sigma}^{(0)}(\mathrm{i}\omega_{1}),\nonumber
\label{52}
\end{eqnarray}
\begin{eqnarray}
\Lambda_{l\sigma l^{\prime}\sigma^{\prime}}(\mathrm{i}\omega)=
\delta_{l l^{\prime}}\delta_{\sigma\sigma^{\prime}}[
m_{l}(\mathrm{i}\omega)+\frac{p^{2}(\mathrm{i}\omega)}{2}\mathcal{G}_{l}^{(0)}(\mathrm{i}\omega)]+
\frac{p^{2}(\mathrm{i}\omega)}{2}
\mathcal{G}_{l}^{(0)}(\mathrm{i}\omega)\mathcal{G}_{l^{\prime}}^{(0)}(\mathrm{i}\omega)
g_{l\sigma l^{\prime}\sigma^{\prime}}(\mathrm{i}\omega),\label{53}
\end{eqnarray}
with the following realizations
\begin{eqnarray}
\Lambda_{11}(\mathrm{i}\omega)&=&
m_{1}(\mathrm{i}\omega)+\frac{p^{2}(\mathrm{i}\omega)}{2}\mathcal{G}_{1}^{(0)}(\mathrm{i}\omega)+
\frac{p^{2}(\mathrm{i}\omega)}{2}[\mathcal{G}_{1}^{(0)}(\mathrm{i}\omega)]^{2}g_{11}(\mathrm{i}\omega), \nonumber \\
\Lambda_{22}(\mathrm{i}\omega)&=&
m_{2}(\mathrm{i}\omega)+\frac{p^{2}(\mathrm{i}\omega)}{2}\mathcal{G}_{2}^{(0)}(\mathrm{i}\omega)+
\frac{p^{2}(\mathrm{i}\omega)}{2}[\mathcal{G}_{2}^{(0)}(\mathrm{i}\omega)]^{2}g_{22}(\mathrm{i}\omega),  \\
\Lambda_{12}(\mathrm{i}\omega)&=&
\frac{p^{2}(\mathrm{i}\omega)}{2}\mathcal{G}_{1}^{(0)}(\mathrm{i}\omega)\mathcal{G}_{2}^{(0)}
(\mathrm{i}\omega)g_{12}(\mathrm{i}\omega).\nonumber\label{54}
\end{eqnarray}%
We take into account the Dyson type equation
\begin{eqnarray}
g_{1 1}(\mathrm{i}\omega)= \frac{\Lambda_{1 1}(\mathrm{i}\omega)-
\mathcal{G}_{2}^{(0)}(\mathrm{i}\omega)(\Lambda_{1
1}(\mathrm{i}\omega)\Lambda_{2 2}(\mathrm{i}\omega) - \Lambda_{1
2}(\mathrm{i}\omega)\Lambda_{2 1}(i\omega))} {d(\mathrm{i}\omega)},
g_{1 2}(\mathrm{i}\omega)=\frac{\Lambda_{1
2}(\mathrm{i}\omega)}{d(\mathrm{i}\omega)},\\
d(\mathrm{i}\omega)=(1-\Lambda_{1
1}(\mathrm{i}\omega)\mathcal{G}_{1}^{(0)}(\mathrm{i}\omega))
(1-\Lambda_{2
2}(\mathrm{i}\omega)\mathcal{G}_{2}^{(0)}(\mathrm{i}\omega)) -
\mathcal{G}_{1}^{(0)}(\mathrm{i}\omega)\mathcal{G}_{2}^{(0)}(\mathrm{i}\omega)\Lambda_{1
2}(\mathrm{i}\omega)\Lambda_{2 1}(\mathrm{i}\omega)).\nonumber
\label{55}
\end{eqnarray}%
We make some generalization by considering function
$m(\mathrm{i}\omega)$ dependent on orbital quantum number $l$ even
if it is really not. The function $g_{22}$ is obtained from equation
(55) by changing indices 1 and 2.

We have found two solutions of equations (54) and (55).

The first of them is
\begin{eqnarray}
\Lambda_{11}(\mathrm{i}\omega)&=&m_{1}(\mathrm{i}\omega),\quad g_{1 1}(\mathrm{i}\omega)=-\frac{1}{\mathcal{G}_{1}^{(0)}(\mathrm{i}\omega)},\nonumber\\
\Lambda_{22}(\mathrm{i}\omega)&=&
\frac{1}{\mathcal{G}_{2}^{(0)}(\mathrm{i}\omega)},\quad
g_{22}(\mathrm{i}\omega)=-\frac{1}{\mathcal{G}_{2}^{(0)}(\mathrm{i}\omega)}
\left(1+
\frac{m_{2}(\mathrm{i}\omega)\mathcal{G}_{2}^{(0)}(\mathrm{i}\omega)-1}{\frac{p^{2}(\mathrm{i}\omega)}{2}(\mathcal{G}_{2}^{(0)}(\mathrm{i}\omega))^{2}}
\right),\\
\Lambda_{12}(\mathrm{i}\omega)&=&\Lambda_{21}(\mathrm{i}\omega)=
\pm\frac{\mathrm{i}p(\mathrm{i}\omega)}{\sqrt{2}},\quad
g_{12}(\mathrm{i}\omega)=\pm\frac{\mathrm{i}\sqrt{2}}{p(\mathrm{i}\omega)
\mathcal{G}_{1}^{(0)}(\mathrm{i}\omega)\mathcal{G}_{2}^{(0)}(\mathrm{i}\omega)},\nonumber\label{56}
\end{eqnarray}%
with the condition that
\begin{equation}
\frac{1}{\mathcal{G}_{1}^{(0)}(\mathrm{i}\omega)}-m_{1}(\mathrm{i}\omega)
=\frac{1}{\mathcal{G}_{2}^{(0)}(\mathrm{i}\omega)}-m_{2}(\mathrm{i}\omega)\label{57}
\end{equation}
The second solution is obtained from (56) by changing indices 1 and 2.

The analytical continuation of obtained solutions in upper
semi-plane gives us the possibility to determine spectral function
of localized electrons
\begin{equation}
\rho_{ll^{\prime}}(E)= -2Im
g_{ll^{\prime}}(E+\mathrm{i}\delta).\label{58}
\end{equation}

For example intraorbital contribution has a form
\begin{eqnarray}
\rho_{11}(E)=-\frac{2Im\mathcal{G}_{1}^{(0)}(E+\mathrm{i}\delta)}{(Re\mathcal{G}_{1}^{(0)})^{2}+(Im\mathcal{G}_{1}^{(0)})^{2}},\label{59}
\end{eqnarray}
where
\begin{eqnarray}
Im\mathcal{G}_{1}^{(0)}(E+\mathrm{i}\delta)=-\pi\rho_{0}(E)|V_{1}|^{2}.\label{60}
\end{eqnarray}

The quantity $\rho_{11}(E)$ differs from zero thanks the existence of the matrix element
of hybridization and of the zero order density of states $\rho_{0}(E)$. For $E=0$ $\rho_{0}(0)$ is positive
and state of the system is metallic.

Interorbital contribution to the phase transition is determined by the
value
\begin{eqnarray}
\rho_{12}(E)=-Img_{12}(E+\mathrm{i}\delta)=\frac{2(E+\triangle
E_{1})(\triangle
E_{2}-E)}{Im\mathcal{G}_{1}^{(0)}(0)Im\mathcal{G}_{2}^{(0)}(0)(\triangle
E_{1}+\triangle E_{2})},\label{61}
\end{eqnarray}
where
\begin{eqnarray}
\triangle E_{1}=E_{2}-E_{9}>0,\quad \triangle E_{2}=E_{12}-E_{9}>0.\nonumber\label{60}
\end{eqnarray}

This quantity is positive for $-\triangle E_{1}<E< \triangle E_{2}$.

For these energy values the state of the system is metallic. The
appearance of spectral weight et the Fermi level is considered as a
definition of Mott transition.

\section{Conclusions}
Diagram approach for investigation the properties of twofold degenerate Anderson
impurity model has been elaborated.

First of all the eigenfunctions and eigenvalues of energy of the localized $d-$ electrons
part of the Hamiltonian have been determined. Their dependence of intra and inter orbital
Coulomb interactions and of Hund rule coupling constant was established.

Perturbation theory around the atomic limit has been developed and Matsubara Green's functions
as in normal and in superconducting states has been defined.

Dyson-type equations for these functions have been established for both states but detailed
solutions were discussed only for normal state supposing additional investigation in the next paper.

Because the main elements of our diagram technique are the irreducible Green's functions we
have undertaken the determination of simplest two-particle irreducible Green's function and determined its dependence of the spin and orbital quantum numbers. This quantity has been determined only in the low temperature
limit.

Having this quantity and summing some class of diagrams we have obtained the $\Lambda_{l\sigma l^{\prime}\sigma^{\prime}}$ correlation function.

We found two solutions for the renormalized Green's functions of the
$d-$ electrons and determined the spectral weight.
\vspace{+5mm}\\
\textbf{Acknowledgment}
\vspace{+5mm}\\
Two of authors (V.M, L.D) would like to express their sincere thanks to Dr. S. Cojocaru for frequent valuable
discussions and comments.

%
%
%
\end{document}